# The Role of the Task Topic in Web Search of Different Task Types

Daniel Hienert[*], Matthew Mitsui[+], Philipp Mayr[*], Chirag Shah[†], Nicholas J. Belkin[†]

[*]GESIS – Leibniz Institute for the Social Sciences
Cologne, Germany
daniel.hienert@gesis.org,
philipp.mayr.gesis.org

[+]Department of Computer Science
Rutgers University
New Brunswick, NJ, USA
mmitsui@cs.rutgers.edu

[†]School of Communication & Information
Rutgers University
New Brunswick, NJ, USA
chirags@rutgers.edu, belkin@rutgers.edu

## ABSTRACT

When users are looking for information on the Web, they show different behavior for different task types, e.g., for fact finding vs. information gathering tasks. For example, related work in this area has investigated how this behavior can be measured and applied to distinguish between easy and difficult tasks. In this work, we look at the searcher's behavior in the domain of journalism for four different task types, and additionally, for two different topics in each task type. Search behavior is measured with a number of session variables and correlated to subjective measures such as task difficulty, task success and the usefulness of documents. We acknowledge prior results in this area that task difficulty is correlated to user effort and that easy and difficult tasks are distinguishable by session variables. However, in this work, we emphasize the role of the *task topic* – in and of itself – over parameters such as the search results and read content pages, dwell times, session variables and subjective measures such as task difficulty or task success. With this knowledge researchers should give more attention to the task topic as an important influence factor for user behavior.

## CCS CONCEPTS

• Information systems~Users and interactive retrieval

## ADDITIONAL KEYWORDS AND PHRASES

User Behavior, Web Search, Task, Topic, Session

## 1 INTRODUCTION

While different models have been proposed for information seeking, in interactive information retrieval (IIR) there is the goal to capture the whole setting with a focus on the interactivity between the *user*, *system*, and *content*. These elementary concepts of information search are, for example, presented by Tsakkonas and Papatheodorou [22] in their triptych framework. Cole et al. [5] apply usefulness as the overall evaluation criterion for each of these components at different levels. The question here is how useful are the systems' results, processes and the delivered content for the leading task and goal, for sub tasks and information seeking strategies (ISS [2]).

The starting moment in this model is the user's *task* which leads the user behavior. This behavior can be described on the system side by a number of session variables, for example by the number of queries or viewed pages within a search session. The *task type* has been identified as one influencing moment of user behavior that can be measured by session variables [11, 20].

However, there are surely more factors which can be found in the triangle system of user, system, and content that influence or can be indicated by session variables. On the user side, there can be factors such as the user's knowledge about the topic and the task, the ability to search efficiently, her or his learning curve or the expectations of the outcome. On the system side, influencing factors can be, e.g., the quality of search engines or the system's support for query terms suggestions or to save and review interesting results. The content side has been a bit unattended in the past of IIR research – that is, the search topic in itself. Also, it is the main source from where users are extracting information from by reading, understanding, and classifying text, images, videos and other information types from Web pages. From a task's view, content can be targeted by the task type (which particular kind of information needs to be extracted?), but also by the task's topic (from which domain, subject area, theme or thing?).

In this paper, we will address this gap by analyzing data from an experiment with four different task types. The experiment's design is insofar specific that each task type is conducted with two different topics. This allows us to examine in particular the role of the task topic whereby the rest of experiment variables (at least on the task-, system-, and user-side) remains stable. We especially examine the relation between subjective user ratings, e.g., for task difficulty, task success and the usefulness of bookmarked pages and session variables such as task time, number of queries or dwell times on read documents. We focus on the investigation which relationships exist and what are the roles of the task topic for user behavior.

## 2 RELATED WORK

### 2.1 The Role of the Task

The idea of a *task* as a motivating moment for the user and as a target variable for the evaluation in interactive information retrieval has gained in importance over the last two decades. Vakkari [23] recites the definition: "A task is an activity to be performed in order to accomplish a goal". Toms [21] gives an outline of the development of the concept "task" and its role in information retrieval. Beside others, an early model of connecting task and search is given by Kekäläinen & Järvelin [10]. They proposed an evaluation model in which the classical lab IR context opened to the information seeking context and work task context. In these contexts, the seeking task and the work task play a major role. Borlund proposed the IIR evaluation model [3] which uses simulated work tasks to simulate information needs and allows the evaluation of IIR systems in a relative controlled environment, but as realistic as possible. Broder [4] suggested an early model for Web search in which he puts the task before the information need. As a first differentiation for task and query types it is differentiated between navigational, informational and transactional queries.

## 2.2 Different Task Types

Differentiating between *task types* helps to study different characteristics of user behavior. For example, Kellar et al. [11] differentiated between the task types fact finding, information gathering, browsing, and transactions. They found that these task types can be distinguished by different characteristics such as task duration, number of viewed pages, the size of queries and the usage of browser functionality. An information gathering task thereby showed to be more complex than a fact finding task. Toms et al. [20] conducted a user study with the three different task types decision making, fact finding and information gathering. Additionally, they explored the effect of two different task structures: (1) parallel, where multiple concepts on the same level are searched and (2) hierarchical, where a single concept is searched, but with multiple characteristics. Li & Belkin [16] propose a faceted task classification system which describe a task on facets such as the source of task, task doer, time, action, product or goal. Cole et al. [6] found behaviors that could distinguish these facets, and additionally adapted this system and added the facet "level of judgment" for their study.

## 2.3 Task Topic and Topic Knowledge

Previous research has also largely explored the relationship between *task topic knowledge* and a searcher's behavior. In evaluation campaigns like TREC, topics are used to describe the scenario for a specific information need which may be described as a mixture of task type and topic (e.g. used in the Core/Web Track [7]). In a more accurate sense, the topic describes the subject (area) of a task [13]. This can be rather a broad domain (e.g. health or e-commerce used in [12]) or a very concrete theme or thing (e.g. a person). Kelly states that the topic represents the focus of the task and that the combination of a specific task and topic forms the information need [13]. On the user side, investigations have been done on how user knowledge may influence search behavior. Thereby it can be distinguished between the broader idea of domain knowledge and the more specific idea of topic knowledge [26]. While domain knowledge describes a general awareness about the broader domain, its content and structure, topic knowledge describes familiarity with the explicit topic (e.g. the concrete theme or thing such as a person, animal or other entities) of the described information need. In general, domain knowledge showed to be influential for the user's search behavior [24, 25]. But also knowledge about the concrete topic showed to have an influence on the searcher's behavior [1, 15, 17].

## 2.4 Subjective Measures and User Behavior

Several works have examined the relationship between subjective measures reported from users and behavioral signals found in log files. Gwizdka and Spence [9] found that variables such as the number of web pages visited or the time spent on each page show correlation to task difficulty for a factual information task on the Web. Gwizdka [8] reports for another experiment that the number of result pages, number of individual pages and number of bookmarks correlate to task difficulty for the two task types fact finding and information gathering. Liu et al. [18] report on the relation between the task type and whole-session in contrast to within-session variables. While whole-session variables describe the session as a whole and can be determined only after a task has finished, within session-variables can be determined at each step of a session and are able to predict task difficulty in real-time. Whole-session variables such as task completion time or number of queries showed a good prediction accuracy to task difficulty. Within-session variables, for example, first dwell time on all SERPs or first dwell time of unique content pages showed a bit lower accuracy for task difficulty prediction. Also, the task type has been shown to influence the prediction level. Kelly et al. [12] conducted an experiment with 20 tasks based on five different complexity levels and four topical domains. They agree that more cognitively complex search tasks require more search activity such as more queries, URL clicks or more time to completion. However, more cognitively complex search tasks were not rated as more difficult by the users and the subjects were equally satisfied with their results across all task types.

## 3 EXPERIMENT

In this section we describe the tasks, the lab study in which these tasks were conducted, and the session variables we use to analyze the participants behavior from the recorded data. In particular we want to address the following research questions:

1. What is the role of the task topic for session variables used to describe user behavior in search sessions?
2. What is the role of the task topic for the relation between subjective user ratings, session variables and dwell times on content pages?

### 3.1 Tasks

Four different tasks were designed, located in the discipline of journalism, which try to capture different search problems in this area. Each of these tasks was conducted with two different topics: (1) "Coelacanth" and (2) "Methane Clathrates and Global Warming". Table 1 presents the different tasks for the topic Coelacanth; the same schema was used for the second topic. Tasks are designed based on the task classification system proposed by [16] and modified in [6]. Table 2 gives an overview of each task type with its task facets. Each participant searched for 2 task types, each task on a different topic. The order of the 2 tasks and 2 topics was additionally flipped, yielding to 16 different configurations.

**Table 1: Search tasks for the topic "Coelacanth"**

| **Assignment 1. Copy Editing (CPE)** |
| --- |
| **Your Assignment:** You are a copy editor at a newspaper and you have only 20 minutes to check the accuracy of the six italicized statements in the excerpt of a piece of news story below. |
| **Your Task:** Please find and save an authoritative page that either confirms or disconfirms each statement. |
| **Assignment 2. Story Pitch (STP)** |
| **Your Assignment:** You are planning to pitch a science story to your editor and need to identify interesting facts about the coelacanth ("see-la-kanth"), a fish that dates from the time of dinosaurs and was thought to be extinct. |
| **Your Task:** Find and save web pages that contain the six most interesting facts about coelacanths and/or research about coelacanths and their preservation. |
| **Assignment 3. Article Development (REL)** |
| **Your assignment:** You are writing an article about coelacanths and conservation efforts. You have found an interesting article about coelacanths but in order to develop your article you need to be able to explain the relationship between key facts you have learned. |
| **Your Task:** In the following there are five italicized passages, find an authoritative web page that explains the relationship between two of the italicized facts. |
| **Assignment 4. Interview Preparation (INT)** |
| **Your Assignment:** You are writing an article that profiles a scientist and their research work. You are preparing to interview Mark Erdmann, a marine biologist, about coelacanths and conservation programs. |
| **Your Task:** Identify and save authoritative web pages for the following: |
| Identify two (living) people who likely can provide some personal stories about Dr. Erdmann and his work. |
| Find the three most interesting facts about Dr. Erdmann's research. |
| Find an interesting potential impact of Dr. Erdmann's work. |

### 3.2 Lab Study

A lab study was conducted with undergraduate students from undergraduate journalism courses having completed at least one course in news writing. The 40 participants had to perform two search tasks (one on each topic), the annotation of bookmarks and search intents and had to fill out a number of questionnaires. Their activity was recorded with a Firefox browser plugin and Morae[1].

The participants started by filling out a demographic questionnaire and by watching a tutorial video of the Firefox plugin. They then filled out the pre-task questionnaire for topic familiarity, assignment experience and assignment difficulty on a 7-point Likert scale (1="not at all" to 7="extremely"). They then had up to 20 minutes time to fulfill the first search task, although they had the option to finish early. Then they there asked to fill out the post-task questionnaire, rating the difficulty of the task, their successfulness in completing the task, whether they had enough time (1="not at all" to 7="extremely"), and whether they understood the task (1="far too little" to 7="more than enough"), on a 7-point Likert scale.

[1] https://www.techsmith.de/morae.html

**Table 2: Task Description and their task facets**

| Task Name | Task Facets [16] | | | |
| --- | --- | --- | --- | --- |
| | Product | Level | Goal | Named Items? |
| Copy Editing | Find facts | Segment | Specific | Yes |
| Story Pitch | Find facts | Segment | Amorphous | No |
| Article Development | Produce ideas | Document | Amorphous | Yes |
| Interview Preparation | Produce ideas | Document | Amorphous | No |

After the search task, the participants were asked to view the video of their task and to annotate the bookmarks and search intentions of their queries. In this process participants were asked to rate the usefulness of each bookmark and their confidence in this rating on a 7-point Likert scale from 1="not at all" to 7="extremely".

The same procedure was then conducted for the second search task. In the exit interview, the users were asked about the experience with the two search tasks. Participants received $30 compensation and $10 for best performance awarded to everyone. The whole study process took about 2 hours per user. In this study we use data from 38 participants, 76 valid search sessions, 20 for Copy Editing (CPE), 18 for Story Pitch (STP), 19 for Article Development (REL), 19 for Interview Preparation (INT) tasks and 38 sessions each for the topic Coelacanth and Methane Clathrates.

### 3.3 Session Variables

To describe the user behavior within a search session, we used a number of session variables following the examples of [12, 18]. We use different categories: (1) Numbers & Frequencies, e.g. action count, (2) ratios, e.g. bookmarks/page visits (3) the overall task time, (4) dwell times, e.g. on content pages, (5) query length, and (6) bookmark dwell times. In [19] different measures for dwell times on content pages were proposed. "Decision time" is the first time within a session the user spends reading on a content page finished by leaving the page e.g. to another tab. "Total dwell time" is the sum of all dwell times the user spends reading a content page. "Total display time" is the whole time span the content page remains open in the browser. In a multi-session experiment Liu & Belkin found that total display time and total dwell time can be a reliable indicator for document usefulness. For category 1 we use the new measures "Number of actions", "Bookmark average first session step" and "First bookmark first session step". Table 3 shows the session variables in detail. Two asterisks at the begin of the variable label indicate a within-session variable.

## 4 RESULTS

In the following we present the results from the pre- and post-task questionnaire, the rated usefulness of bookmarks and session variables per topic and task type. All values in the presented tables are color-coded. This allows comparison between each task, but also within one task type. Over all tables we use red for higher values and green for lower values. The idea is to instantly see patterns for certain task types or topics.

**Table 3: Session Variables, \*\*=Within-session variable**

| | Variable | Definition |
|---|---|---|
| Numbers & Frequencies | **# Actions** | Total number of user interactions including queries, page visits, adding/selecting/closing tabs, save/delete bookmarks, copy&paste text |
| | # Unique queries | Number of unique user queries |
| | # SERP visits | Number of SERP visits |
| | # Unique page visits | Number of unique page visits |
| | # Page visits | Number of total page visits |
| | # Unique bookmarks | Number of unique bookmarks |
| | # Bookmarks | Number of total bookmarks |
| | **\*\*Bookmark average first session step** | The average first session step over all bookmarked pages. |
| | **First bookmark first session step** | The session step for the first bookmarked page. |
| | # Searches without page visits | Number of searches without page visits |
| Ratios | \*\*Unique pages/unique searches | Ratio of unique pages per search |
| | \*\*Pages/unique searches | Ratio of pages per search |
| | Bookmarks/page visits | Ratio of bookmarks per page visit |
| | Unique Bookmarks/Unique page visits | Ratio of bookmarks per unique page visit |
| | Unique Bookmarks/Unique queries | Ratio of unique bookmarks per unique page visit |
| | Bookmarks/Unique queries | Ratio of bookmarks per unique search |
| TT | Task Time | Time for the whole task |
| Dwell times | Total time on content pages | Total time on all content pages |
| | \*\*Average time on content pages | Average time on content pages |
| | Total time on SERPs | Total time on all SERPs |
| | \*\*Average time on SERPs | Average time on SERPs |
| Query length | Total query length | Total query length in characters |
| | Average query length | Average query length in characters |
| Bookmark dwell times | \*\*Bookmark decision time | Total decision time on all bookmarks |
| | \*\*Non-bookmark decision time | Total decision time on pages not bookmarked |
| | \*\*Bookmark total dwell time | Total dwell time on all bookmarks |
| | \*\*Non bookmark total dwell time | Total dwell time on pages not bookmarked |
| | \*\*Bookmark total display time | Total display time on all bookmarks |
| | \*\*Non bookmark total display time | Total display time on pages not bookmarked |

When dividing data by both task type and topic, there are less than 12 participants per group, making it difficult to perform statistical tests for some variables. Hence, some subsequent analyses where we divide both by task and topic, significance testing is absent. However, we provide significance tests where we only divide by one factor such as task or topic, and also for within-session variables (explained in Section 5).

## 4.1 Pre-task Subjective Measures

In the pre-task questionnaire, subjects were asked for their topic familiarity, experience, and the perceived difficulty level after reading the task assignment. Table 4 shows the average results per task type and topic. *Topic familiarity* with the topic Coelacanths was on average rated with 1.3 ("not at all familiar") and for Methane Clathrates with 2.4 ("low familiarity"). This is a margin of 1.1 towards the topic Methane Clathrates. (significantly different with Mann-Whitney with p<0.0001). The *assignment experience* within a task type was rather stable with the rating "slight experience" with 2.80 for Copy Editing, 3.39 for Story Pitch, 2.53 for Article Development and 3.47 for Interview preparation. (from the same population with Kruskal-Wallis test). The perceived *difficulty level* was relatively stable for the two topics within each task type, but diverging between task types with 3.75 ("somewhat difficult") for Copy Editing, 3.39 ("slightly difficult") for Story Pitch, 4.37 ("somewhat difficult") for Article development and 3.89 ("somewhat difficult") for Interview preparation (not significantly different). We can observe that the topic "Coelacanth" seems to be less familiar to subjects than "Methane Clathrates" over all task types. The task "Article Development" with the topic "Coelacanth" was rated most challenging based on topic familiarity, assignment experience, and perceived difficulty.

## 4.2 Post-task Subjective Measures

After conducting the task, users were asked to fill out a post-task questionnaire for the difficulty and success of the task, the availability of enough time and the understanding of the assignment. Table 5 shows the average results per task type and topic. For *post difficulty*, inverse to the pre-task statements, the topic "Coelacanth" seems to be easier than "Methane Clathrates" over all task types with 2.39 ("low difficulty") for Coelacanths and 2.89 ("slight difficulty") for Methane Clathrates. Post difficulty also diverges between task types with 2.60 ("slight") for Copy Editing, 1.50 ("low") for Story Pitch, 3.16 ("slight") for article development and 3.26 for Interview preparation ("slight"). *Success* was rated better for the topic Coelacanth with 5.61 ("very successful") than for Methane Clathrates ("moderately successful"). The task type Story Pitch was rated with 6.22 ("very successful") and the task Copy Editing with 5.35 ("moderately"), Article development and Interview preparation both with 4.84 ("moderately successful"). *Enough time* was felt moderately for the task type Story Pitch with 4.67 and especially for the topic Coelacanth with 4.90 ("more than enough"). The rest of task types show values around 4 ("enough"). For *comprehension*, we can see high values for the task type Story Pitch with 6.33 ("understood very well") and lower values for Interview preparation with 5.42 ("understood moderately well").

Overall, the task type "Story Pitch" seems to be the easiest task type based on the average measures of difficulty, success, enough time and comprehension. This is followed by "Copy Editing" with a bit lower values. Then comes "Article Development" and most difficult to do was "Interview Preparation".

**Table 4: Mean pre-task subjective measures**

| Task type | Topic | Topic familiarity | Assignment experience | Pre-Difficulty |
|---|---|---|---|---|
| Copy Editing | Coelacanth | 1.56 | 2.56 | 3.56 |
| | Methane | 2.64 | 3.00 | 3.91 |
| Story Pitch | Coelacanth | 1.20 | 3.40 | 3.50 |
| | Methane | 2.13 | 3.38 | 3.25 |
| Article development | Coelacanth | 1.11 | 2.33 | 4.78 |
| | Methane | 2.40 | 2.70 | 4.00 |
| Interview preparation | Coelacanth | 1.20 | 3.40 | 3.80 |
| | Methane | 2.22 | 3.56 | 4.00 |

**Table 5: Mean post-task subjective measures**

| Task type | Topic | Post-Difficulty | Success | Enough time | Comprehension |
|---|---|---|---|---|---|
| Copy Editing | Coelacanth | 2.11 | 5.56 | 4.56 | 6.11 |
| | Methane | 3.00 | 5.18 | 3.64 | 5.73 |
| Story Pitch | Coelacanth | 1.20 | 6.50 | 4.90 | 6.30 |
| | Methane | 1.88 | 5.88 | 4.38 | 6.38 |
| Article development | Coelacanth | 3.11 | 5.11 | 4.00 | 5.44 |
| | Methane | 3.20 | 4.60 | 4.10 | 5.70 |
| Interview preparation | Coelacanth | 3.20 | 5.20 | 4.20 | 5.80 |
| | Methane | 3.33 | 4.44 | 3.89 | 5.00 |

**Table 6: Session Variables by topic and task type (time values are in seconds; ** = Within-session variable)**

| Task type | Topic | # Actions | # Unique queries | # SERP visits | # Unique page visits | # Page visits | # Unique bookmarks | # Bookmarks | **Bookmark average first session step | First bookmark first session step | # Searches without page visits | **Unique pages/searches | **Pages/searches | Bookmarks/page visits | Unique Bookmarks/Unique page visits | Unique Bookmarks/Unique queries | Bookmarks/Unique queries | Task time | Total time on content pages | **Average time on content pages | Total time on SERPs | **Average time on SERPs | Total query length | Average query length |
|---|---|---|---|---|---|---|---|---|---|---|---|---|---|---|---|---|---|---|---|---|---|---|---|---|
| Copy Editing | Coelacanth | 183.78 | 9.00 | 25.89 | 12.56 | 32.89 | 6.11 | 6.67 | 91.02 | 21.89 | 2.44 | 2.91 | 6.53 | 0.28 | 0.57 | 0.94 | 1.01 | 792.56 | 372.33 | 11.32 | 125.11 | 4.83 P=0.028 | 366.33 | 43.22 |
| | Methane | 214.36 | 9.91 | 24.55 | 15.00 | 40.55 | 6.55 | 6.91 | 107.62 | 20.27 | 1.64 | 3.11 | 6.57 | 0.21 | 0.47 | 0.71 | 0.75 | 1003.73 | 507.09 | 12.51 | 164.73 | 6.71 | 560.36 | 61.64 |
| Story Pitch | Coelacanth | 80.80 | 3.10 | 13.50 | 10.40 | 17.70 | 5.20 | 5.30 | 135.34 P<0.001 | 11.60 | 0.20 | 4.65 P=0.003 | 10.06 P=0.019 | 0.36 | 0.52 | 2.08 | 2.18 | 549.10 | 375.90 | 21.24 | 90.30 | 6.69 P=0.046 | 105.20 | 32.80 |
| | Methane | 135.38 | 7.75 | 22.75 | 15.13 | 30.50 | 6.25 | 6.75 | 165.72 | 15.38 | 2.75 | 3.60 | 6.87 | 0.34 | 0.50 | 1.15 | 1.23 | 814.50 | 508.00 | 16.66 | 183.38 | 8.06 | 429.25 | 50.38 |
| Article development | Coelacanth | 174.56 | 7.67 | 23.89 | 14.78 | 36.67 | 4.89 | 5.00 | 94.76 | 47.44 | 1.44 | 3.70 P=0.010 | 7.90 P=0.044 | 0.15 | 0.38 | 0.64 | 0.64 | 992.56 | 518.11 | 14.13 | 149.22 | 6.25 P=0.002 | 337.00 | 40.44 |
| | Methane | 189.50 | 9.00 | 34.20 | 14.60 | 28.90 | 6.40 | 6.50 | 105.06 | 26.50 | 3.50 | 2.86 | 7.01 | 0.29 | 0.49 | 0.90 | 0.90 | 864.00 | 394.60 | 13.65 | 186.40 | 5.45 | 376.10 | 43.50 |
| Interview preparation | Coelacanth | 211.40 | 10.30 | 30.50 | 17.60 | 58.10 | 6.90 | 7.00 | 102.21 | 22.20 | 2.10 | 3.68 | 8.52 | 0.15 | 0.44 | 0.87 | 0.89 | 908.70 | 582.60 | 10.03 | 190.60 | 6.25 P=0.002 | 418.20 | 35.20 |
| | Methane | 161.89 | 6.33 | 21.33 | 15.44 | 40.89 | 7.00 | 7.44 | 93.48 | 24.22 | 1.11 | 4.16 | 9.82 | 0.23 | 0.52 | 1.98 | 2.03 | 861.67 | 531.56 | 13.00 | 170.11 | 7.97 | 276.22 | 37.44 |

**Table 7: Mean usefulness of bookmarks and confidence in usefulness ratings**

| Task type | Topic | Usefulness of bookmarks | Confidence in bookmark rating |
|---|---|---|---|
| Copy Editing | Coelacanth | 6.04 | 6.19 |
| | Methane | 5.59 | 5.85 |
| Story Pitch | Coelacanth | 6.06 | 5.87 |
| | Methane | 5.71 | 5.72 |
| Article development | Coelacanth | 5.42 | 5.36 |
| | Methane | 5.49 | 5.57 |
| Interview preparation | Coelacanth | 5.78 | 6.03 |
| | Methane | 5.53 | 5.48 |

**Table 8: Mean dwell times for bookmark and non-bookmark content pages**

| Task type | Topic | Decision Time | | Total dwell time | | Total display time | |
|---|---|---|---|---|---|---|---|
| | | Bookmark | Non-bookmark | Bookmark | Non-bookmark | Bookmark | Non-bookmark |
| Copy Editing | Coelacanth | 20.60 | 6.31 P<0.0001 | 48.60 | 11.09 P=0.005 | 149.93 | 54.88 |
| | Methane | 17.96 | 9.36 | 46.18 | 15.12 | 167.71 | 82.12 |
| Story Pitch | Coelacanth | 44.13 P=0.039 | 7.62 | 57.43 P=0.043 | 11.42 | 99.83 P=0.024 | 52.77 P=0.043 |
| | Methane | 27.69 | 8.02 | 50.37 | 10.99 | 204.28 | 56.11 |
| Article development | Coelacanth | 28.71 | 8.38 P=0.002 | 61.84 | 12.73 P<0.001 | 282.42 | 74.85 |
| | Methane | 25.43 | 6.88 | 35.66 | 13.66 | 84.20 | 65.30 |
| Interview preparation | Coelacanth | 22.91 | 5.13 | 58.80 | 9.00 P<0.001 | 294.24 | 60.95 P=0.028 |
| | Methane | 29.87 | 6.37 | 55.22 | 11.51 | 138.21 | 89.15 |

### 4.3 Average Values of Session Variables

In Table 6 we show the average results for each session variable for each task ordered by task type and topic. This table gives an overview which values can be expected for different task types. We will not go into detail of every single value. However, as a first impression, for the lowest rated task in difficulty "Story Pitch – Coelacanth" with 1.20, frequencies such as the number of actions, over the number of searches without page visits to average query length and task time are the lowest, ratios are mostly the highest. Total time on content pages and SERPs is low, but average times are high. The other way around, the task "Interview preparation" – with the same topic "Coelacanth" shows high values for frequencies, low ratios, and high numbers for total time on content pages and SERPs.

### 4.4 Usefulness of Bookmarks

In a separate session after conducting the task subjects then rated the usefulness of individual bookmarks and their confidence in these ratings. Table 7 shows the average rating per task type and topic. Highest rating for the usefulness of bookmarks is for "Story Pitch" and the topic "Coelacanth" with 6.06 ("very useful"), lowest for "Article Development" and "Coelacanth" with 5.42 ("moderately useful"). The differences between the task types were rather low with "Story Pitch" 5.90, "Copy Editing" 5.79, "Interview Preparation" with 5.66 (all three "very useful") and Article Development with 5.46 ("moderately useful").

### 4.5 Dwell Times on Content Pages

We also computed the average dwell times for bookmark and non-bookmark content pages per task type and topic based on the measures *decision time*, *total dwell time* and *total display time* proposed by [19], Table 8 shows the results. Decision times for bookmarks are from 17.96s to 44.13s, for non-bookmarks pretty stable from 5.31sec to 9.36s. Total dwell times for bookmarks are from 35.66s to 61.84s, for non-bookmarks from 9.00s to 15.12s. Total display times for bookmarks are from 84.20s to 294.24s, for non-bookmarks from 52.77s to 89.15s. This means for each dwell time measure bookmarked content pages have a significant higher dwell time than non-bookmarks. This statement is valid for all task topics and types. However, decision times seem to be diverse across task type and topic.

## 5   Analysis I

Values and colors in Table 6 give a first impression that session variables are also dependent on the task topic, not only on the task type. For example, the number of actions in the task type Story Pitch is different for the topic Coelacanth with 80.80 actions and for Methane Clathrates with 135.38 actions. Or, the task time for Copy Editing is different for Coelacanth with 792.56s to Methane Clathrates with 1003.73s. Therefore in this section we analyze the role of the task topic for session variables and dwell times. We also acknowledge that there is a significant difference in topic familiarity. Moreover, about 26% of participants searching for Methane Clathrates reported a high familiarity (4 or above) while only about 3% reported high familiarity for Coelacanths. We therefore also report findings for only users with low familiarity, analyzing both the full pool of sessions and also those where participants had low topic familiarity.

### 5.1   Session Variables

As mentioned before, we wanted to compare the mean values for a significant statistical difference between topics of one task type. However, dividing the data set first by task type and then to topics gives very small groups of only up to eleven subjects for each topic. Also because of possible high standard deviations it is hard to find statistical significance. Here, more subjects and data for each topic would be needed. However, some session variables give more than one data point per session and user. In [18] they are called within-session variables because these variables can be gathered also in the middle of a user session. These are mainly 'number of (unique) content pages per query' and 'first (mean) dwell time on content pages or SERPs'.

Unique pages per search and pages per search (n=2,732) showed significant differences for the topic in the task type Story Pitch (*p=0.003*, *p=0.019*) and Article Development (*p=0.010*, *p=0.044*) with a Mann–Whitney test with *alpha=0.05*. We also checked the differences for dwell time on content pages and SERPs and found significant differences in topics for all task types for time on SERPs (*p=0.028*, *p=0.046*, *p=0.002*, *p=0.002*). Each statistical significant difference between topics of one task type is marked with a bold line and p-values on the left side of the table cell in the Tables 6 and 8.

In this experiment, we additionally used the session variable 'bookmark first session step'. We compared the values for all bookmarks (n=490) between topics within a task type. This showed a statistical difference for the task type Story Pitch with *p<0.001*. Altogether, 9 from 20 values of within-session variables show significant differences between topics.

To analyze the influence of topic familiarity, we additionally examined the session variables for all sessions with a topic familiarity of 1-3 ("overall low", n=65 sessions). The mean values remain stable with only slight changes, also, all significant differences between topics remain valid.

### 5.2   Dwell times

Dwell times for content pages are a good indicators and give better statistical results, because every user's view on a content page is a new data point. Table 8 show the mean values for the different dwell times. Here again we seek for statistical differences between topics in one task type.

For bookmarked pages (n=490) we found significant different dwell times for Story Pitch – Decision Time (*p=0.039*) and Total display time (*p=0.024*), Article Development – Total dwell time (*p=0.002*) and Total display time (*p<0.001*) and Interview Preparation – Total display time (*p<0.001*).

For normal content pages (non-bookmark, n=2,181) we found different dwell times for topics for Copy Editing – Decision Time (*p<0.0001*) and Total dwell time (*p=0.005*), Story Pitch – Total dwell time (*p=0.043*) and Total display time (*p=0.043*), and Interview Preparation – Total display time (*p=0.028*). This means, different dwell times differ dependent on the topic. Here, 10 of 24 values in dwell times show significant differences between topics.

Here again, we analyze the influence of topic familiarity by examining the dwell times for all sessions with a topic familiarity of 1-3. The mean values remain stable with only slight changes. The significance for dwell time differences between topics remained stable for all reported ones without two: Story Pitch – Bookmark decision time and Bookmark total display time.

## 6   Analysis II

In this section we conduct a number of correlation analyses to find relationships between pre- and post-task measures, session variables, usefulness ratings and dwell times, also dependent on the task type and topic.

### 6.1   Pre-task Measures

First, we did a correlation analysis from pre-task measures to post-task measures and to session variables (see Tables 9a-c).

*6.1.1 Topic Familiarity.* For topic familiarity we found only weak overall correlations. For the topic Coelacanth there is a weak correlation to post-difficulty (0.348) which is significantly different to Methane Clathrates by a margin of 0.460. Dividing the data set by task type we found no correlations from topic familiarity to post subjective measures, but depending on the task type to different session variables: for Copy Editing to Average query length (0.562) and for Story Pitch to Number of unique bookmarks (0.514) and Number of bookmarks (0.585).

*6.1.2 Assignment Experience.* For assignment experience there are also only weak overall negative correlations to other subjective and session variables. The topic Methane Clathrates shows correlations to post-difficulty, average time on SERPs and bookmark decision time which Coelacanth does not. Coelacanth shows a moderate correlation to Comprehension which Methane Clathrates does not.

**Table 9:** Spearman correlation for pre-task subjective measures with at least significant correlation in one column (in bold different from zero with a significance level of alpha=0.05). Correlation values between Coelacanth and Methane Clathrates significantly different with Fisher's r to z transformation and p<0.05 in bold.

(a) Topic Familiarity

| Variable | All | Coe. | Met. | Diff. |
|---|---|---|---|---|
| Post-Difficulty | 0.114 | **0.348** | -0.112 | **0.460** |
| # Bookmarks | **0.266** | 0.264 | 0.280 | 0.016 |
| Average query length | **0.397** | 0.123 | 0.287 | 0.164 |
| Non bookmark total display time | **0.321** | 0.038 | 0.026 | 0.012 |

(b) Assignment Experience

| Variable | All | Coe. | Met. | Diff. |
|---|---|---|---|---|
| Pre-Difficulty | **-0.287** | -0.207 | **-0.353** | 0.146 |
| Post-Difficulty | **-0.334** | -0.227 | **-0.464** | 0.237 |
| Success | **0.268** | 0.220 | **0.353** | 0.133 |
| Comprehension | **0.347** | **0.404** | 0.282 | 0.122 |
| Confidence in bookmark rating | **0.226** | 0.198 | 0.250 | 0.052 |
| Average time on content pages | 0.130 | 0.109 | **0.162** | 0.053 |
| Average time on SERPs | 0.206 | 0.079 | **0.340** | 0.260 |
| Bookmark decision time | 0.015 | -0.173 | **0.262** | **0.435** |
| Bookmark total dwell time | 0.095 | 0.026 | **0.282** | 0.256 |

(c) Pre-Difficulty

| Variable | All | Coe. | Met. | Diff. |
|---|---|---|---|---|
| Assignment experience | **-0.287** | -0.207 | **-0.353** | 0.146 |
| Post-Difficulty | **0.310** | 0.303 | **0.388** | 0.085 |
| Success | -0.225 | -0.197 | **-0.338** | 0.141 |
| Comprehension | **-0.286** | **-0.423** | -0.208 | 0.215 |
| Unique Bookmarks/Unique queries | -0.128 | **-0.336** | 0.143 | **0.480** |
| Bookmarks/Unique queries | -0.128 | **-0.349** | 0.144 | **0.493** |
| Average time on SERPs | 0.086 | **0.340** | -0.140 | **0.481** |

**Table 10:** Spearman correlation for post-task subjective measures with at least significant correlation in one column (in bold different from zero with a significance level of alpha=0.05). Correlation values between Coelacanth and Methane Clathrates significantly different with Fisher's r to z transformation and p<0.05 in bold.

(a) Post-Difficulty

| Variable | All | Coe. | Met. | Diff. |
|---|---|---|---|---|
| Topic familiarity | 0.114 | **0.348** | -0.112 | **0.460** |
| Assignment experience | **-0.334** | -0.227 | **-0.464** | 0.237 |
| Pre-Difficulty | **0.310** | 0.303 | **0.388** | 0.085 |
| Success | **-0.689** | **-0.664** | **-0.676** | 0.012 |
| Enough time | **-0.633** | **-0.573** | **-0.631** | 0.058 |
| Comprehension | **-0.637** | **-0.710** | **-0.575** | 0.135 |
| Usefulness of bookmarks | **-0.469** | **-0.455** | **-0.419** | 0.035 |
| Confidence in bookmark rating | **-0.414** | -0.314 | **-0.455** | 0.141 |
| # Actions | **0.467** | **0.485** | **0.414** | 0.070 |
| # Unique queries | **0.323** | **0.421** | 0.148 | 0.273 |
| # SERP visits | **0.385** | **0.426** | 0.302 | 0.125 |
| # Unique page visits | **0.356** | 0.237 | **0.461** | 0.224 |
| # Page visits | **0.383** | **0.387** | **0.402** | 0.015 |
| # Unique bookmarks | 0.234 | 0.180 | 0.265 | 0.085 |
| Bookmark average first session step | **0.456** | **0.425** | **0.452** | 0.027 |
| First bookmark first session step | **0.229** | **0.331** | 0.115 | 0.216 |
| # Searches without page visits | **0.233** | 0.285 | 0.092 | 0.194 |
| Unique pages/unique searches | -0.078 | **-0.341** | 0.245 | **0.586** |
| Bookmarks/page visits | **-0.299** | **-0.378** | -0.240 | 0.138 |
| Unique Bookmarks/Unique page visits | **-0.278** | -0.225 | **-0.334** | 0.109 |
| Unique Bookmarks/Unique queries | **-0.257** | **-0.429** | -0.042 | **0.387** |
| Bookmarks/Unique queries | **-0.241** | **-0.424** | -0.031 | **0.394** |
| Task time | **0.556** | **0.500** | **0.548** | 0.048 |
| Total time on content pages | **0.358** | 0.276 | **0.360** | 0.084 |
| Total query length | **0.291** | **0.503** | 0.003 | **0.501** |
| Bookmark total display time | **0.270** | **0.412** | 0.245 | 0.167 |
| Non bookmark total display time | **0.228** | 0.237 | 0.199 | 0.038 |

(b) Success

| Variable | All | Coe. | Met. | Diff. |
|---|---|---|---|---|
| Assignment experience | **0.268** | 0.220 | **0.353** | 0.133 |
| Pre-Difficulty | -0.225 | -0.197 | **-0.338** | 0.141 |
| Post-Difficulty | **-0.689** | **-0.664** | **-0.676** | 0.012 |
| Enough time | **0.625** | **0.701** | **0.547** | 0.154 |
| Comprehension | **0.537** | **0.695** | **0.422** | 0.272 |
| Usefulness of bookmarks | **0.560** | **0.549** | **0.564** | 0.015 |
| Confidence in bookmark rating | **0.568** | **0.519** | **0.557** | 0.038 |
| # Actions | **-0.318** | **-0.387** | -0.216 | 0.171 |
| # Unique queries | -0.202 | **-0.354** | -0.004 | 0.350 |
| # SERP visits | **-0.329** | **-0.452** | -0.192 | 0.260 |
| Bookmark average first session step | **-0.340** | **-0.408** | -0.230 | 0.178 |
| First bookmark first session step | **-0.280** | **-0.447** | -0.112 | 0.335 |
| Unique pages/unique searches | 0.081 | **0.349** | -0.242 | **0.591** |
| Unique Bookmarks/Unique queries | 0.170 | **0.369** | -0.056 | **0.425**[+] |
| Bookmarks/Unique queries | 0.170 | **0.357** | -0.034 | **0.391**[+] |
| Task time | **-0.456** | **-0.501** | **-0.385** | 0.115 |
| Total time on content pages | **-0.270** | **-0.279** | -0.215 | 0.064 |
| Total query length | -0.189 | **-0.366** | 0.050 | **0.416**[+] |

(c) Enough time

| Variable | All | Coe. | Met. | Diff. |
|---|---|---|---|---|
| Post-Difficulty | **-0.633** | **-0.573** | **-0.631** | 0.058 |
| Success | **0.625** | **0.701** | **0.547** | 0.154 |
| Comprehension | **0.434** | **0.513** | **0.359** | 0.154 |
| Usefulness of bookmarks | **0.520** | **0.551** | **0.461** | 0.090 |
| Confidence in bookmark rating | **0.399** | **0.348** | **0.378** | 0.030 |
| # Actions | **-0.543** | **-0.529** | **-0.538** | 0.008 |
| # Unique queries | **-0.436** | **-0.513** | **-0.381** | 0.132 |
| # SERP visits | **-0.462** | **-0.467** | **-0.470** | 0.003 |
| # Unique page visits | **-0.356** | -0.176 | **-0.455** | 0.279 |
| # Page visits | **-0.431** | **-0.458** | **-0.434** | 0.024 |
| Bookmark average first session step | **-0.511** | **-0.485** | **-0.498** | 0.013 |
| First bookmark first session step | -0.197 | **-0.378** | -0.022 | 0.356 |
| # Searches without page visits | -0.264 | **-0.313** | -0.178 | 0.135 |
| Unique pages/unique searches | 0.203 | **0.467** | 0.000 | **0.467** |
| Bookmarks/page visits | **0.371** | **0.420** | **0.375** | 0.045 |
| Unique Bookmarks/Unique page visits | **0.347** | 0.285 | **0.395** | 0.110 |
| Unique Bookmarks/Unique queries | **0.372** | **0.521** | 0.261 | 0.260 |
| Bookmarks/Unique queries | **0.362** | **0.499** | 0.271 | 0.228 |
| Task time | **-0.705** | **-0.616** | **-0.759** | 0.143 |
| Total time on content pages | **-0.535** | **-0.452** | **-0.577** | 0.124 |
| Total time on SERPs | **-0.271** | -0.219 | -0.220 | 0.001 |
| Total query length | **-0.450** | **-0.547** | **-0.334** | 0.213 |
| Average query length | **-0.348** | **-0.338** | -0.206 | 0.132 |
| Bookmark total dwell time | -0.259 | -0.212 | **-0.404** | 0.192 |
| Non bookmark total dwell time | 0.128 | **0.361** | 0.010 | 0.351 |
| Bookmark total display time | **-0.400** | **-0.625** | **-0.365** | 0.260 |
| Non bookmark total display time | **-0.284** | -0.103 | **-0.483** | **0.380**[+] |

(d) Comprehension

| Variable | All | Coe. | Met. | Diff. |
|---|---|---|---|---|
| Assignment experience | **0.347** | **0.404** | 0.282 | 0.122 |
| Pre-Difficulty | **-0.286** | **-0.423** | -0.208 | 0.215 |
| Post-Difficulty | **-0.637** | **-0.710** | **-0.575** | 0.135 |
| Success | **0.537** | **0.695** | **0.422** | **0.272**[+] |
| Enough time | **0.434** | **0.513** | **0.359** | 0.154 |
| Usefulness of bookmarks | **0.399** | **0.469** | 0.313 | 0.156 |
| Confidence in bookmark rating | **0.338** | **0.417** | 0.236 | 0.181 |
| # Page visits | **-0.232** | -0.228 | -0.248 | 0.020 |
| Task time | **-0.310** | **-0.418** | -0.200 | 0.218 |
| Total time on content pages | **-0.280** | **-0.320** | -0.190 | 0.130 |
| Non bookmark total dwell time | 0.080 | **0.358** | -0.228 | **0.586** |
| Bookmark total display time | -0.218 | **-0.352** | -0.178 | 0.174 |

(e) Usefulness of bookmarks

| Variable | All | Coe. | Met. | Diff. |
|---|---|---|---|---|
| Post-Difficulty | **-0.469** | **-0.455** | **-0.419** | 0.035 |
| Success | **0.560** | **0.549** | **0.564** | 0.015 |
| Enough time | **0.520** | **0.551** | **0.461** | 0.090 |
| Comprehension | **0.399** | **0.469** | 0.313 | 0.156 |
| Confidence in bookmark rating | **0.662** | **0.748** | **0.554** | 0.194 |
| # Actions | **-0.321** | **-0.380** | -0.234 | 0.147 |
| # SERP visits | **-0.250** | **-0.344** | -0.167 | 0.177 |
| # Page visits | **-0.282** | **-0.367** | -0.223 | 0.144 |
| Bookmark average first session step | **-0.244** | **-0.308** | -0.143 | 0.165 |
| Bookmarks/page visits | **0.256** | **0.300** | 0.285 | 0.014 |
| Task time | **-0.466** | **-0.610** | **-0.317** | 0.293 |
| Total time on content pages | **-0.376** | **-0.493** | -0.248 | 0.245 |
| Average time on SERPs | **0.323** | 0.282 | **0.476** | 0.194 |
| Total query length | -0.239 | **-0.321** | -0.134 | 0.186 |
| Average query length | -0.227 | -0.024 | -0.230 | 0.206 |
| Bookmark total dwell time | **-0.323** | -0.290 | **-0.455** | 0.165 |
| Bookmark total display time | **-0.396** | **-0.544** | **-0.396** | 0.148 |

*6.1.3 Pre-Difficulty.* Also for pre-difficulty we found only weak overall correlations. Coelacanth shows negative correlations to ratios unique bookmarks/unique queries and bookmarks/unique queries which Methane Clathrates does not with a margin of nearly 0.5.

## 6.2 Post-task Measures

Following this line, we conducted a correlation analysis from subjective measures of the post questionnaire and from the usefulness ratings of bookmarks to session variables. Tables 10a-e show the summarized results.

*6.2.1 Post-Difficulty.* For post-difficulty there are strong overall correlations to other post-questionnaire measures such as to success (-0.689), enough time (-0.633) and comprehension (0.637). There are also moderate correlations from post-difficulty to the usefulness of bookmarks (-0.469) and confidence in the bookmarks (-0.414). From task difficulty to session variables we have found moderate correlations to the number of actions (0.467), bookmark average first session step (0.456) and task time (0.556). Additionally, we can find a number of weak correlations to other session variables.

For different *topics* the correlations to subjective measures, number of actions, bookmark average first session step, and task time are stable. However, other session variables such as bookmarks/unique queries or total query length differ.

For different *task types* we find for Copy Editing a correlation to Ratio unique pages/unique searches (0.561). For Story Pitch to task time (0.723), total time on content pages (0.572) and total time on SERPs (0.516). For Article Development there are no correlations and for Interview preparation to searches without page visits (0.460), task time (0.694) and total time on SERPs (0.579).

*6.2.2 Success.* Success is also strongly correlated to other of post-questionnaire measures such as enough time (0.625) and comprehension (0.537). There is also a high correlation to usefulness of bookmarks (0.560) and confidence in bookmark rating (0.568). We found a moderate negative correlation from success to task time (-0.456) and some weak correlation to other session variables. The topics differ on session variables such as unique bookmarks/unique queries.

*6.2.3 Enough Time.* Enough time correlates moderately to strongly to a number of other subjective measures and session variables. The two topics here are relatively stable, only between unique pages/unique searches there is a margin up to 0.467.

*6.2.4 Comprehension.* Comprehension shows moderate to strong correlations for enough time and post-difficulty. Here, a lot of correlations are moderately and significant for Coelacanth, but not for Methane Clathrates.

*6.2.5 Usefulness of Bookmarks.* Usefulness of bookmarks is weakly to moderately correlated to the post-questionnaire measures post-difficulty (-0.469), success (0.560), enough time (0.520), and comprehension (0.399) and strongly correlated to the confidence in the rating (0.662). For *session variables*, there is a moderate correlation to task time (-0.466) and several other weak correlations.

If we divide the data set by *topic*, we find for Coelacanth a strong negative correlation of -0.610 to task time, and moderate negative correlations -0.544 to bookmark total display time and of -0.493 to total time on content pages. For the topic Methane Clathrates correlations to task time and total time on content show a weaker correlation with a difference around 0.3.

If we divide the data by *task type*, we can find for Copy Editing correlations to average query length (-0.500). For Story Pitch to task time (-0.577), total time on content pages (-0.608) and bookmark total display time (-0.567). For the type Article Development we find correlations to task time (-0.546) and average query length (-0.509). No correlations were found for the task type Interview preparation.

We also tested with sessions of topic familiarity 1-3. Then the significant difference test between topics failed for values marked with a plus in the tables 10b Success, 10c Enough Time, and 10d Comprehension.

## 7 DISCUSSION

### 7.1 Pre-task Subjective Measures

The factor of *topic familiarity* in this experiment had an overall weak effect on the number of bookmarks and a nearly moderate effect on query length. This means user behavior is influenced slightly by making more bookmarks and moderately by entering longer queries for those who felt they had more familiarity with a topic. For the topic Coelacanth, topic familiarity showed a weak influence to post difficulty with 0.348, for Methane Clathrates it did not. Also, the different task types showed no influence from topic familiarity to post-difficulty. We additionally checked session variables and dwell times for the influence of topic familiarity. This showed only minor changes in mean values and most significant differences between topics remain intact. For the *experience in the assignment* there is a weak negative correlation to pre- and post-difficulty around -0.28 to -0.33. For the *pre-difficulty* measure we found a weak correlation to post difficulty (0.31) and a weak negative correlation to comprehension (-0.28). No overall session variables were influenced by the pre-task difficulty. This means, in this experiment the perceived difficulty before the task has only a slight influence on perceived difficulty after the task and on task success. All in all, pre-task measures here have only a weak effect on task behavior and post-task ratings. Only topic familiarity to query length has a nearly moderate effect.

### 7.2 Post-task Subjective Measures

*7.2.1 Task Difficulty and User Effort.* The correlation analysis for post-task measures showed that *task difficulty* is correlated to a number of session features which in general measure the *user effort* to conduct the task. The variable task time is a general feature which can represent user effort for a task and shows a solid correlation with 0.556. A novel tested measure in this study is the number of overall actions which shows a stable correlation of 0.467 and describes the number of all interactions the user does. More fine-grained features representing user effort are number of SERP visits and number of page visits with still moderate correlation around 0.38. Some other session variables representing user effort show still weak correlations such as total time on content pages (0.35) and total query length around 0.29. Most of these features have also been found to correlate with task difficulty in related work [e.g. 8, 12, 18]. Action count, task time and bookmark average first session step showed stable correlations also for both task topics. However, these correlations cannot be found in all task types. It seems intuitive that more user effort results in the subjective impression that the task is more difficult. However, the correlation is not so strong for every task type that there is a direct one-to-one relationship. So, also other factors seem to influence the subjective task difficulty level.

*7.2.2 Task Difficulty and Task Success.* In this experiment, we found a strong negative correlation of -0.689 between task difficulty and task success. This means, the more difficult a task felt, the less successful it was rated. Again, this sounds intuitive, and there seems to be a strong overall relationship between difficulty and success. Session variables which correlate with both concepts are task time, first bookmark first session step and bookmark average first session step. Task time for both has a moderate correlation (0.556 vs. -0.456) and can describe the overall effort as described above. Two new features Bookmark average first session step has a correlation to task difficulty with 0.456 and to success with -0.340. These feature describe when in the session (e.g. sooner or later described in action steps) on average the bookmarks are saved. They can describe in a simple manner when first results for a task are found.

*7.2.3 Task Difficulty and Usefulness of Content Pages.* We also found a moderate correlation of -0.469 between task difficulty and of 0.560 between success and the usefulness of bookmarked web pages. This means, the more useful the bookmarked web pages were seen, the less difficult and more successful the task was rated. This is a clear indication that the usefulness of the bookmarked content (the task's results) has an influence on the task difficulty and success. This is surely an intuitive notion; however, the usefulness of the content has not yet been taken into account so far as to measure the success and the difficulty of a task. For sure this aspect has been discussed on a model basis [e.g. 22, 5] and has been researched for decades on the basis of the relevance of the content to a user query. But, the usefulness of the content in relation to a task measured over a whole user session is still a different issue.

*7.2.4 Dwell Times.* Related work has found that dwell times on content pages can be used to predict document usefulness under consideration of the task type and also the specific user [14]. In [19] decision time, total dwell time and total display time were examined in two different task types: a dependent task and a parallel task. While total dwell time and total display time were good predictors for usefulness in each individual task, decision time was not. The authors argued that in the parallel task the sub tasks only changed in their topic and users could reuse some useful documents.

*7.2.4.1 Aspect Threshold.* Also in this experiment different types of dwell times show significant differences between bookmarked (and usefully rated) pages and those which were not bookmarked. Decision time and total dwell times are relatively stable for non-bookmarking pages over all tasks and topics. So, for decision time there is a range from 5.13s to 9.36s for non-bookmarking and from 17.96s to 44.13s for bookmarked pages. A certain threshold, e.g. of 14s, here can surely predict those pages which will be bookmarked by the user. The same is true and even enforced for total dwell time: there is a range from 9.00s to 15.12s for non-bookmarking pages and from 35.66s to 61.84s for bookmarked pages. A threshold of e.g. 20sec could surely predict those pages which will be bookmarked. The picture is not that clear for total display time. Here, time span are overlapping between the span of bookmarked and non-bookmarked pages: from 52.77s to 89.15s for non-bookmarked and from 84.20s to 294.24s for bookmarked. And, for each task the times are significantly different for bookmarked and non-bookmarked pages.

*7.2.4.2 Aspect Usefulness of Bookmarks.* In this experiment, over all tasks we found weak to moderate negative correlations between usefulness of bookmarks and dwell times, e.g. for bookmark total display time. This is in contrast to related work [e.g. 19], where longer dwell times correlate with higher usefulness ratings. However, other correlations seem to be dependent on the topic. For the topic Coelacanth we find a strong negative correlation to task time and a moderate to total time on content pages which cannot be found for the topic Methane Clathrates. We have to mention that in this experiment we have only usefulness ratings for bookmarked pages, not for every content page. This might influence the correlation analysis for these session variables. However, other session variables such as total time on content pages are available for every content page. We also tested for the relationship between usefulness of bookmarks and different task types. Here, we also find different results. For the task type Story Pitch there is a moderate to strong correlation for task time and time on content pages which could not be found for the other task types.

## 7.3 Task Type and Task Topic

The mean values for different session variables in Table 8 and correlations in Table 10 give a first indication that user behavior is not only dependent on the task type, but also on the task topic.

In the section 'Analysis I' we found a number of session variables that show significant differences between the two topics in one task type, e.g. bookmark first session step or pages/search. Especially, different dwell time measures show significant differences between topics. There are two reasons for that: (a) in some cases (e.g. for Story Pitch with the task level 'Document segment') we found that high decision times originate from individual web pages with a lot of text on it. So, users need up to several minutes for the extraction of the relevant information for the task. (b) In other cases, users spend more time on average on all content pages.

In the section 'Analysis II' we looked for overall correlations between subjective measures and session variables. This can be set in contrast to correlations found by dividing the dataset by task type or topic. For example, *topic familiarity* only showed a weak correlation for the topic Coelacanth, but not for Methane Clathrates and not for different task types. For correlations to *task difficulty* the session variables action count, task time and average first session step showed stable correlations for both topics, but not for each task type. Other correlations to post-difficulty are dependent on the topic. For the *usefulness of bookmarks* a number of correlation can be found for the topic Coelacanth (-0.610 to task time, -0.544 to bookmark total display time and -0.493 to total time on content pages) which are weaker for Methane Clathrates.

## 8 CONCLUSION

In this work we analyzed data from an experiment with four different task types and additionally two different topics for each task type. User behavior was measured with session variables and dwell times. We mainly conduct two analyses: (I) a comparison of mean values and (II) a correlation analysis from subjective user rating such as task difficulty, task success and usefulness of bookmarked pages to session variables. From the analysis and the discussion we conclude the following points:

- *Topic familiarity* in this experiment overall only played a minor role because both topics were fairly unfamiliar to subjects. But topic familiarity was dependent on the task topic.
- *Task difficulty* is moderately correlated to *user effort* and can be measured with a number of session variables such as task time, number of actions, or more specifically with features such as number of SERP visits or number of content pages. The correlation between user effort and task difficulty seems to be dependent on the task type and topic.
- Session variables measuring user behavior are also dependent on the task type and task topic.
- *Task success* and *task difficulty* are strongly negatively correlated, and task success can be measured with session variables such as task time and with session variables dependent on the topic.
- *Task success* and the *usefulness of bookmarks* interpreted as the task's result are nearly strong related. This means the content's usefulness plays an important role for the task's success.
- *Usefulness of bookmarks* is weakly to moderately correlated to certain dwell times and dependent on the task type and topic.
- A *threshold* can be used to distinguish between *useful (bookmarked) pages* and other content pages. Decision time and total dwell time can be used as within-session variables independent of the task type and topic.
- *Decision time* on web pages can be dependent on the text size on the page and how easy it is to extract the relevant information for the user. This is dependent on the task type and topic.

Therefore the *task type*, but also the *task's topic* has an important influence on user behavior. The task type influences how users are searching; the task topic influences what results are presented by the search engine. The search results influence dwell times, and nearly all session variables. This influences at the end the perceived task success and difficulty.

If researchers are using only one topic in their task description, this can massively influence the results in a free Web search task. A good solution for this issue has been applied by Kelly et al. in their study [12] who used four domains (health, commerce, entertainment, science & technology) and different topics tailored to study participants in the sense of Borlund's Simulated Work Task [3].

**Acknowledgements** This work was partly funded by the DFG grant no. MA 3964/5-1 and by the NSF grant no. IIS-1423239.